\documentclass[letterpaper,12pt]{article}
\usepackage{tabularx} % extra features for tabular environment
\usepackage[utf8]{inputenc}
\usepackage{amsmath}  % improve math presentation
\usepackage{graphicx} % takes care of graphic including machinery
\usepackage[margin=1in,letterpaper]{geometry} % decreases margins
\usepackage[titletoc]{appendix}
\usepackage{float}
\usepackage[final]{hyperref}
\usepackage[style=phys]{biblatex}
\usepackage[capitalize,noabbrev]{cleveref}
\usepackage{physics}
\hypersetup{
	colorlinks=true,       % false: boxed links; true: colored links
	linkcolor=blue,        % color of internal links
	citecolor=blue,        % color of links to bibliography
	filecolor=magenta,     % color of file links
	urlcolor=blue         
}
\DeclareUnicodeCharacter{0301}{e}
\DeclareUnicodeCharacter{0300}{e}
\begin{document}

\title{Bound states of two-photon Rabi model at the collapse point}

\author{Ching Kwan Chan\thanks{Department of Physics, The Chinese University of Hong Kong }}

\date{June 9, 2020}
\maketitle

\begin{abstract}
This paper presents a proof of the existence of novel bound states of the two-photon quantum Rabi model at the collapse point. The two-photon Rabi model is interesting not only for its important role on non-linear light-matter interaction, but also for the exhibition of many-energy-levels degenerating process called the "spectral collapse". The squeezing property of the two-photon annihilation and creation operators is the origin for this phenomenon which is well studied without the energy-slitting term $\omega_0$. However, many numerical studies have pointed out that with the presence of $\omega_0$ , some low-level isolated states exist while other high energy states collapse to $E=-\frac{\omega}{2}$, which known as incomplete spectral collapse. From the eigenvalue equation in real space, pair of second order differential equations, which are similarly to the Schrodinger equation, are derived at the collapse point. These differential equations provide explanation to the existence of isolated bound states below $E=-\frac{\omega}{2}$ with the presence of the spin slitting $\omega_0$ and better numerical method to generate those bound states.
\end{abstract}

\maketitle

\section{\label{sec:level1}Introduction}

The interaction between matter and photon system remarks a central problem in recent development of quantum technology. The quantum Rabi model(QRM) is the simplest model describing the interaction between spin and bosonic system($\hbar=1$) \cite{1,2}.
\begin{equation}
\hat{H}=\omega a^\dag a+\frac{1}{2}\omega_0\sigma_z+\epsilon(a^\dag+ a)\sigma_x
\end{equation}
where the $\omega$ is the frequency of the bosonic mode, $\omega_0$ is the spin-$\frac{1}{2}$ energy-slitting and $\epsilon$ is the coupling between two sub-systems. With the recent quantum technology development, different regimes of the coupling can be reached, especially the ultra-strong coupling regime, where $\epsilon$ is comparable to $\omega$ and the deep-strong regime where $\epsilon>\omega$ \cite{18,19,20,21,27}. In those strong coupling regimes, rotating wave approximation(RWA) and perturbation method do not work anymore \cite{39}. Previously unexplored regions can now be investigated and theoretical treatments are needed for a better understanding of light-matter interaction. Some interesting applications are expected using those strong coupling regimes \cite{19}. On the other hand, Braak successfully derived the analytical exact spectrum of QRM, which called the G-function method \cite{17}. This theoretical breakthrough has simulated many studies on QRM and generalization of QRM. \cite{28,29,30,31,32}\\

The two-photon Rabi model is a non-linear generalization of QRM, which has the following Hamiltonian($\hbar=1$),
\begin{equation}
\hat{H}=\omega a^\dag a+\frac{1}{2}\omega_0\sigma_z+\epsilon({a^\dag}^2+ a^2)\sigma_x
\label{2pqrm}
\end{equation}
 This model was experimentally realised by advanced quantum technology like trapped ion \cite{3,7}, circuit quantum electrodynamics\cite{4,6} and other quantum platforms \cite{33,34}. Similar to QRM, the strong coupling regime of two-photon Rabi model can be achieved on different quantum platforms. Thus interest in studying the two-photon Rabi model has grown in recent years.\\

This model is exactly solvable for only two cases: $\omega_0=0$, where the solutions are the squeezed number states, or when $\epsilon=0$, where the solutions are the number states. However, the full Hamiltonian cannot be solved analytically \cite{8}. The most interesting phenomenon of two-photon Rabi model happens when $\epsilon\rightarrow\omega/2$, the energy levels start to collapse towards $E=-\omega/2$, called the spectral collapse \cite{7,9,10,11}. This can be explained using the case of $\omega_0=0$. In real space representation, the Hamiltonian is just a quantum harmonic oscillator. When $\epsilon\rightarrow\omega/2$, the quadratic potential vanishes and becomes a free particle system, resulting in a continuum system with no bound state \cite{9}. However, numerous studies pointed out that bound states exist at the collapse point when $\omega_0\neq0$, which refers as the incomplete spectral collapse \cite{5,13}. This effect has not been clearly explained.\\

The most common way to study the two-photon Rabi model is using the numerical diagonalization \cite{9,5}, which requires lots of computational power, especially towards the collapse point as higher number states are getting involved. Recently, a numerical method based on the spectral function and continued fraction has been studied \cite{13}, which allows a high truncation number with reasonable computational time. However, this method just focuses on the spectrum of the system. The numerical method is difficult for studying the strong coupling regime and fails beyond the collapse point. It is because the wavefunction is widely spread and even no longer bounded, so it cannot be represented by number states. Fake converging states have been reported for $\epsilon>\omega/2$ if truncation number is not large enough \cite{9,13}. The truncated wavefunction is stabilized by the local potential but globally it is unstable \cite{5}. Therefore, the numerical observation does not provide strong evidence for the existence of bound states at the collapse point. Analytical treatment is needed to prevent ambiguity in numerical methods.\\

The theoretical development of two-photon Rabi model revolves mainly around variational approximation \cite{5} and the spectrum of the system. The recent variational study of two-photon Rabi model used the concept of polaron. The polaron picture provides a helpful understanding for the wavefunction with different coupling strength. Using this method, the spin slitting term produces an effective trapping potential which allows bound states exist while the quadratic potential vanishes at the collapse point \cite{5}. This method gives a qualitative description about the incomplete spectral collapse. However, the complete analysis of the wavefunction at the collapse point is largely unexplored. Another way to study two-photon Rabi model is using G-function method \cite{14,15,16,23,17}. The roots of the G-function would be the eigenenergies of the two-photon Rabi model. The G-function can be calculated using the analytic function in Fock-Barmman space or by Bogoliubov operators approach. By studying the poles structure of G-function, the structure of the spectral collapse can be found out. However, because of the nature of spectral collapse, the poles are highly degenerated near collapse point. This method is also failed to explain the incomplete spectral collapse \cite{10}.\\

A research has been done on the Rabi-Stark model which is another generalization of QRM with Stark-like term \cite{26}. A real space approach has been used to explain the spectral accumulation occurs at critical coupling. The system has an effective quadratic potential which depends on the system energy. Self-consistent solution can be obtained analytically under this method to explain the existence of the spectral accumulation phenomenon.\\

With many tools to study the two-photon Rabi model, however, the isolated bound states are not yet clearly explained. Both current theoretical and numerical methods have difficulties in dealing with the collapse point and beyond. Here we provide a simple solution to explain and prove the existence of the bound states at the collapse point using similar approach as the Rabi-Stark model. The scheme of this paper is as follows. First, in the next section, we derive the fourth-order differential equations of the two-photon Rabi model and study the corresponding simplified version at the collapse point. Then, the concepts of effective potential and effective energy are introduced for this Schrodinger-like differential equation. Finally, numerical evidences are provided to show the consistency with previous studies. 

\section{\label{sec:level1}Fourth-order differential equation in real space representation}

\subsection{\label{sec:level2}Hamiltonian in \texorpdfstring{$\sigma_x$}{basis} basis}
The original Hamiltonian of the two-photon Rabi model is given in Eq.(\ref{2pqrm}). A convenient way to rewrite the Hamiltonian in spin-boson representation is applying the unitary transform $e^{-\frac{i\pi}{4}\sigma_y}$\cite{10,5}. The transformed Hamiltonian becomes
\begin{equation}
    \hat{H}=\omega a^\dag a+\frac{1}{2}\omega_0\sigma_x+\epsilon({a^\dag}^2+ a^2)\sigma_z
\end{equation}
In this form we have actually changed the basis from spin along $z$-axis to $x$-axis. The $\omega_0$ term is acting as a tunneling term\cite{5}. The corresponding wavefunction is in form of two functions
\begin{equation}
    \ket{\Psi}= \begin{pmatrix}
           \psi_{+}(x) \\
           \psi_{-}(x)
         \end{pmatrix}
\label{wf}
\end{equation}
The real space operators can be defined from the creation and annihilation operators $a^\dag$ $a$ as follows($m=1$),
\begin{equation}
    \hat{x}=\frac{1}{\sqrt{2\omega}}(a+a^\dag), \quad \hat{p}=-i\frac{d}{dx}=\frac{1}{i}\sqrt{\frac{\omega}{2}}(a-a^\dag)
\end{equation}
Using the real space representation, the eigenvalue equation can be written as follow
\begin{equation}
    \hat{H}\ket{\Psi}=\omega \bigg(\frac{1}{2}\hat{p}^2+\frac{1}{2}\omega^2x^2-\frac{1}{2}\bigg)\ket{\Psi}+\frac{1}{2}\omega_0\sigma_x\ket{\Psi}+\frac{2\epsilon}{\omega}\bigg(-\frac{1}{2}\hat{p}^2+\frac{1}{2}\omega^2x^2\bigg)\sigma_z\ket{\Psi}=E\ket{\Psi}
\label{eve}
\end{equation}

Substituting $\ket{\Psi}$ in Eq.\eqref{wf} into Eq.\eqref{eve}, the eigenvalue equation becomes two coupled second-order differential equations for spin-up and spin-down real space functions. 
\begin{equation}
    \begin{cases}
\omega_0\psi_+(x) =(1+\frac{2\epsilon}{\omega}) \psi''_-(x)-(1-\frac{2\epsilon}{\omega})\omega^2x^2\psi_-(x)+(2E+\omega)\psi_-(x)\\
\omega_0\psi_-(x) =(1-\frac{2\epsilon}{\omega}) \psi''_+(x)-(1+\frac{2\epsilon}{\omega})\omega^2x^2\psi_+(x)+(2E+\omega)\psi_+ (x)\label{eq:1}
\end{cases}
\end{equation}
where $\psi'(x)$ is the derivative of $\psi(x)$ with respect to $x$. \\

If $\omega_0=0$ and $\epsilon=\omega/2$, the equations becomes two decoupled free-particle equations in real space and momentum space. This explained the origin of spectral collapse of the two-photon Rabi model. In the previous polaron method study, the tunneling term is treated as an effective potential $V_{eff}\approx\omega_0 \psi_\pm/ \psi_\mp$. The additional effective potential enhanced the trapping property for different coupling. Therefore without the quadratic potential at the collapse point, the system is possible to have bound state \cite{5}. This method provides qualitative description and physical insight, however the variational state is not the true eigenstate and it may not capture the true property at the collapse point. More rigorous treatment are needed to provide a clearer description at the collapse point. \\

Eq.(\ref{eq:1}) can further be decoupled to form two fourth-order equations. 
\begin{align}
\bigg(1-&\frac{4\epsilon^2}{\omega^2}\bigg) \psi''''_\pm(x)-\bigg[2\bigg(1+\frac{4\epsilon^2}{\omega^2}\bigg)\omega^2x^2-4E-2\omega\bigg]\psi''_\pm(x) 
-4\bigg(1\pm\frac{2\epsilon}{\omega}\bigg)^2\omega^2x\psi'_\pm(x)\nonumber\\
&+\bigg[\bigg(1-\frac{4\epsilon^2}{\omega^2}\bigg)\omega^4x^4-(4E+2\omega)\omega^2x^2+(2E+\omega)^2-2\bigg(1\pm\frac{2\epsilon}{\omega}\bigg)^2\omega^2-\omega_0^2\bigg]\psi_\pm(x)=0
 \label{eq:wideeq}
\end{align}

These two differential equations can be easily checked with other numerical result. The equations are valid for every regime as no approximation has been made. Although these fourth-order differential equations linearise the effect of $\omega_0$, there is no better way to deal with Eq.(\ref{eq:wideeq}) other than numerical method. The complete analytical study of these fourth order differential equations is really difficult.\\

Noted that the two wavefunctions $\psi_\pm(x)$ actually form a Fourier transform pair, namely that they are related by Fourier transform. This relation can be explained by the equivalence of Fourier transform and $\sigma_z$ operator, which would be discussed in detail in Appendix. This property ensures if one of the wavefunction is bounded, another wavefunction must be bounded too because the inner product is invariant under Fourier transform. Therefore we can safely investigate just one of the equations to study when bound states can be formed.\\

\subsection{\label{sec:level2}Second-order differential equation at \texorpdfstring{$\epsilon=\frac{\omega}{2}$}{} and asymptotic behaviour }
At the collapse point $\epsilon=\omega/2$, the fourth order differential equations can be much simplified into second order differential equations similar to the work on Rabi-Stark model \cite{26}. Then these equations can be handled more easily using the similarity to Schrodinger equation. 
\begin{align}
    -&\psi''_+(x)-\frac{16\omega^2x}{4\omega^2x^2-4E-2\omega}\psi'_+(x)+\frac{-(4E+2\omega)\omega^2x^2+(2E+\omega)^2-8\omega^2-\omega_0^2}{4\omega^2x^2-4E-2\omega}\psi_+(x)=0
    \label{psip}
\end{align}

\begin{equation}
    -\psi''_-(x)+\frac{-(4E+2\omega)\omega^2x^2+(2E+\omega)^2-\omega_0^2}{4\omega^2x^2-4E-2\omega}\psi_-(x)=0
    \label{psin}
\end{equation}
Eq.(\ref{psip}) and (\ref{psin}) are second order linear differential equations with rational function coefficients. As stated in the previous section, $\psi_+(x)$ and $\psi_-(x)$ are related by Fourier transform. Therefore we can focus on studying one of them without worrying the other is bounded or not. Here $\psi_-(x)$ is chosen as the target. The boundedness property of the wavefunction is determined by the asymptotic behaviour at the infinity. In fact, the asymptotic behaviour is just a simple differential equation with constant coefficient,
\begin{equation}
    \psi''_-(x)+(E+\frac{\omega}{2})\psi(x)\sim 0, \quad x\rightarrow\pm\infty
\end{equation}
\begin{equation}
    \psi_-(x)\sim \exp(\mp\sqrt{-E-\frac{\omega}{2}}x), \quad x\rightarrow\pm\infty
\end{equation}
One thing should be noticed that $E$ implicitly depends on $\omega_0$. The  bounded states can only exist when the wavefunction is normalizable, which means the argument inside the square root has to be positive. This explained why isolated states exist only for $E<-\omega/2$. To simplify the following calculation, we introduce a redefined energy 
\begin{equation}
    \widetilde{E}=E+\frac{\omega}{2}
\end{equation}
So the condition of bound states become $\widetilde{E}<0$ and the collapse differential equation for $\psi_-(x)$ becomes
\begin{equation}
    -\psi''_-(x)+\frac{-4\widetilde{E}\omega^2x^2+4\widetilde{E}^2-\omega_0^2}{4\omega^2x^2-4\widetilde{E}}\psi_-(x)=0
\label{cpeqn}
\end{equation}
\subsection{\label{sec:level2}Effective potential and energy}
Eq.\eqref{cpeqn} looks very much the same as the Schrodinger equation but differs by a factor of 2.
\begin{equation}
    -\frac{1}{2}\Psi''+V(x)\Psi(x)=E\Psi(x)
\end{equation}
Therefore we introduce the concepts of effective potential and effective energy for Eq.\eqref{cpeqn}.
\begin{equation}
    V_{eff}(x,\widetilde{E},\omega_0,\omega)=\frac{1}{2}\frac{-4\widetilde{E}\omega^2x^2+4\widetilde{E}^2-\omega_0^2}{4\omega^2x^2-4\widetilde{E}} \quad \quad E_{eff}=0
\end{equation}
However, it is not a traditional Schrodinger-type problem, which varies the eigenenergy with fixed potential to search for bound state. Here we actually vary the effective potential with parameters $\widetilde{E}$ and $\omega_0$ with fixed effective energy $E_{eff}=0$ to search for bound state. Unfortunately, this second order differential equation cannot be solved analytically, but we can still get some insights by plotting out the effective potential.\\

\begin{figure}[H]
\includegraphics[width=7cm]{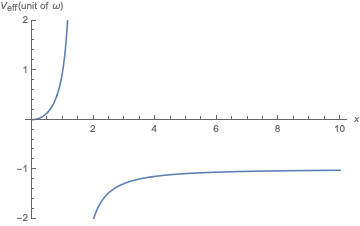}
\centering
\caption{The figure shows the effective potential for $\widetilde{E}>0$, setting $\omega_0=4$, $\widetilde{E}=2.5$ and $\omega=1$ as example. The potential asymptotically approaches constant$=-\widetilde{E}$ at infinity.  }
\label{veff1}
\end{figure}

\begin{figure}[H]
\includegraphics[width=7cm]{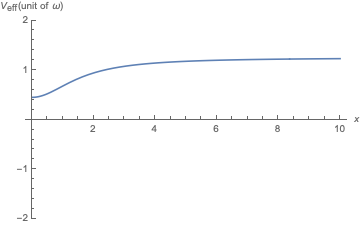}
\centering
\caption{The figure shows the effective potential for $\widetilde{E}<-\omega_0/2$, setting $\omega_0=4$, $\widetilde{E}=-2.5$ and $\omega=1$. The potential is attractive but greater than 0 for all x.  }
\label{veff2}
\end{figure}

\begin{figure}[H]
\includegraphics[width=10cm]{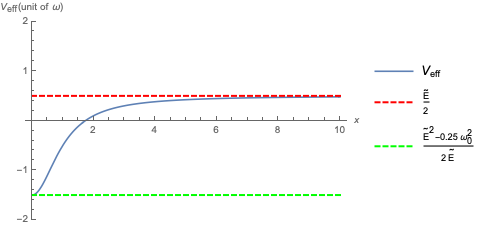}
\centering
\caption{The figure shows the effective potential for $-\omega_0/2<\widetilde{E}<0$, setting $\omega_0=4$, $\widetilde{E}=-1$ and $\omega=1$. The potential is bowl-like shaped. The upper and lower bounds of the effective potential have shown in the figure. The depth of the potential is $\frac{\omega_0^2}{8\widetilde{E}}$. Therefore, the number of bound state mainly depends on the value of $\omega_0$}
\label{veff3}
\end{figure}

With the concept of the effective potential, the problem becomes whether the effective potential can support bound state with zero energy. The \cref{veff1,veff2,veff3} show the different cases of the effective potential, setting $\omega_0=4$ and $\omega=1$ as examples. First in \cref{veff1}, the $\widetilde{E}>0$ case, the potential flattens below 0 at large $x$. Therefore it is impossible to support bound states with zero energy. Then, for the case of $\widetilde{E}<-w_0/2$ as shown in \cref{veff2}, the whole potential is above 0. As the operator $\hat{p}^2$ is positive definite, it is again impossible to have bound state with zero energy for this case. Therefore, only $-\omega_0/2<\widetilde{E}<0$ case, the bowl-like potential well can support bound states. In practice,  $\omega_0$ is the only controllable external parameter and $\widetilde{E}$ is a measurable. Therefore, the depth of the effective potential is controlled by $\omega_0$. The deeper the potential well, it can support more bound states. We can understand this problem as quantum finite potential well. There will be at least one bounded state for $\omega_0\neq0$. Just like finite quantum well, even the depth is infinitesimally small, there would be at least one bound state. The number of state depends on the depth and the width of the well. Similarly, $\omega_0$ controls the depth of the effective potential, so the number of the bound state depends on the value of $\omega_0$. \\

On the other hand, the two-photon Rabi model is originally not a simple one-dimensional problem because of the spin-slitting term. In the real space representation of Eq.(\ref{2pqrm}), $\sigma_z\rightarrow(-1)^{\hat{n}/2}=\exp{\frac{i\pi}{4}(\hat{p}^2+\hat{x}^2)}$ is a non-local potential. With this non-local potential, the node theorem cannot be applied and degeneracy can be found in the two-photon Rabi model \cite{35,36}. But the problem reduced back into one-dimensional quantum problem at the collapse point. Therefore, we can know the excitation number of the state by just simply counting its node.\\

From the collapse equation, we can confirm some numerical observations. First, the collapse point $\epsilon=\omega/2$ is independent of presence of $\omega_0$. No matter what is the value of $\omega_0$, the energy levels above $E=-\omega/2$ will become a continuum when $\epsilon$ reaches $\omega/2$. Second, the bound state can only exist under $E<-\omega/2$. Third, the energy of ground state is bounded below for $E_g>-\omega_0/2$. Final, there are more bound states when $\omega_0$ is increased. These observations can be explained through the collapse point equations.  

\section{\label{sec:level1}Validation with numerical result}
In this section, we are going to examine the consistency with numerical and analytical results. Here we employ the diagonalization method as the basic method. The disadvantage of this method is the eigenstates require high truncation number for certain accuracy. However, if we only consider the low-lying energy levels, it is possible to have a lower truncation number. Especially when $\omega_0$ is large, which makes the effective potential deeper and localises the wavefunction. However, if we consider bound state for small $\omega_0$ or bound state with energy $\widetilde{E}\sim0^-$, the wavefunction is widely spread and requires higher truncation number.\\

\subsection{\label{sec:level2}Asymptotic behavior of numerical result}

One characteristic of the wavefunction at the collapse point is presence of the exponential decay tail. For $\epsilon<\omega/2$, the wavefunction decays like a Gaussian. By simply taking the logarithm of the wavefunction, we can see the quadratic nature before the collapse point and linear nature at the collapse point.
\begin{figure}[H]
\includegraphics[width=15cm]{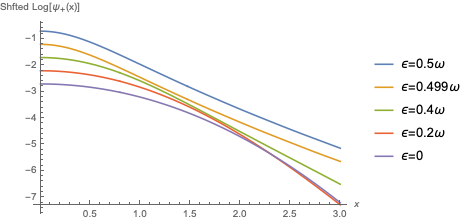}
\centering
\caption{The figure shows the logarithm of wavefunctions $Log[\psi_+(x)]$ for $\omega_0=2$ and $\epsilon=0, 0.2, 0.4, 0.499, 0.5\omega$. The plots are shifted vertically for demonstration purpose. The tails of the logarithm change from quadratic to linear when $\epsilon$ approaching $0.5\omega$. The logarithm of wavefunction at the collapse point is asymptotically linear for large $x$.  }
\end{figure}

The exponential tail shows the quadratic potential vanished at the collapse point. The bound state is solely held by well-like potential. This confirms the prediction of effective potential by polaron method \cite{5}. The asymptotic behavior of the wavefunction is no longer Gaussian but exponentially delay. This may be reason why multipolaron method, which uses sum of Gaussian functions, works well while single polaron, which is impossible to capture the asymptotic behavior, failed at the collapse point.

\subsection{\label{sec:level2}Numerical integration of wavefunction at the collapse point}

\begin{figure}[H]
\includegraphics[width=14cm]{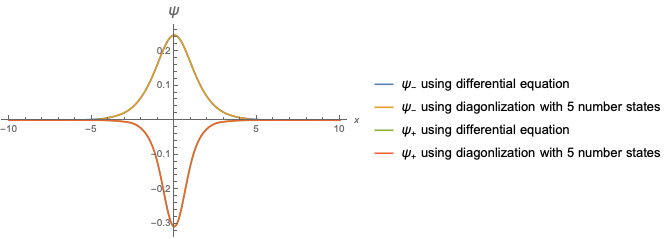}
\centering
\caption{The figure shows the ground state wavefunctions $\psi_\pm(x)$ with $\omega_0=4$ generated by solving differential equations and diagonalization at the collapse point. For the diagonalization, the cutoff number=2000, and only first 5 number states have used for approximation. The wavefunctions generated by two different methods are matched very well and merged. }
\label{w_0=4cp}
\end{figure}

\begin{figure}[H]
\includegraphics[width=14cm]{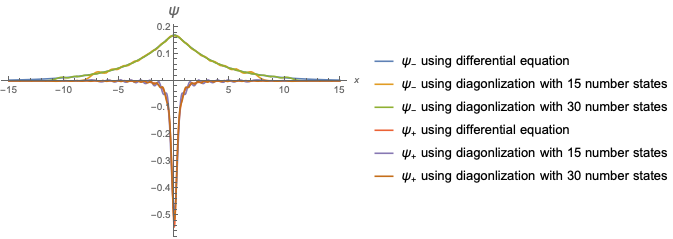}
\centering
\caption{The figure shows the ground state wavefunctions $\psi_\pm(x)$ with $\omega_0=0.5$ generated by integrating differential equations and diagonalization at the collapse point. For the diagonalization, the cutoff number=2000, and first 15 and 30 number states have included to approximate the $\psi_\pm(x)$. The $\psi_\pm(x)$ require more number states as $\omega_0$ is smaller. }
\label{w_0=05cp}
\end{figure}

It is easy to numerically integrate the second-order differential equation at the collapse point. Noticed that the ratio of $\psi_\pm(x)$ is important and can be found using Eq.\eqref{eq:1}, we can find that
\begin{equation}
    \psi_+(0)=\frac{\omega_0}{2E+\omega}\psi_-(0)\quad, \quad \psi'_+(0)=\frac{\omega_0}{2E+\omega}\psi'_-(0) 
\end{equation}
Because of the symmetry of the two-photon Rabi model, $\psi_\pm(x)$ would be odd when it is odd excited state and vice versa. So with odd or even initial conditions, it can generate the non-normalized wavefunction. Then, we can use numerical integration to normalize the wavefunctions with following conidtions
\begin{equation}
\int^\infty_{-\infty}\psi^2_+(x)dx+\int^\infty_{-\infty}\psi^2_-(x)dx=1
\end{equation}
After normalization, we can compare the ground wavefunctions generated by differential equations and diagonlization. When large $\omega_0$ was used, they match perfectly with just few number states used as shown in \cref{w_0=4cp}. It is because $\omega_0$ creates a deep effective potential, and localizes the wavefunctions. However, when $\omega_0$ is small, $\psi_-(x)$ is wildly spread and $\psi_+(x)$ is squeezed, more number states have to be included to approximate $\psi_\pm(x)$ as shown in \cref{w_0=05cp}. This increases the difficulty of diagonalization. Therefore, in the case of wildly spread or squeezed $\psi_\pm(x)$, numerical integration of the collapse point differential equations is much easier and accurate.\\

\subsection{\label{sec:level2}Creating bound state with increasing \texorpdfstring{$\omega_0$}{}}

From the collapse point differential equations, the two-photon Rabi model has a well like effective potential at the collapse point, which means number of bound states increases with $\omega_0$. One easier way of finding new bound state is using numerical diagonalization to check the energy of the state. If the energy of the state gets smaller than $-\omega/2$ while increasing $\omega_0$, then it would be a new created bound state. However, it is very difficult to numerically generate a state with $E\sim-\omega/2$, as they are widely spread. Using this method, the first excited state is formed at $\omega_0\approx1.095\omega$, and the second excited state is formed at $\omega_0\approx1.145\omega$. Because of the numerical difficulty, here we demonstrate the first excited state at $\omega_0=1.2\omega$ with $E=-0.500385\omega$ and second excited state at $\omega_0=1.3\omega$ with $E=-0.500200\omega$.

\begin{figure}[H]
\includegraphics[width=8cm]{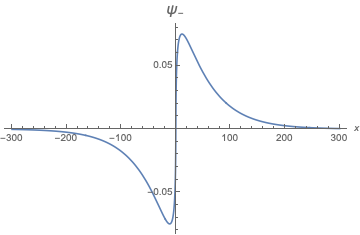}
\includegraphics[width=8cm]{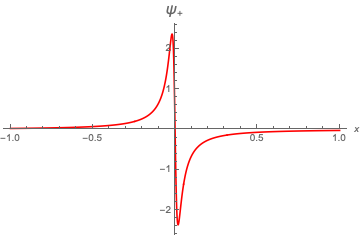}
\centering
\caption{The figures show the normalized first excited state $\psi_\pm(x)$ generated by numerical integration of the collapse differential equations with $\omega_0=1.2$, $E=-0.500385$ and $\omega=1$. The $\psi_-(x)$ is wildly spread while $\psi_+(x)$ is well localized. One interesting thing is that $\int^\infty_{-\infty}\psi^2_+(x)dx=\int^\infty_{-\infty}\psi^2_-(x)dx$, which means the probability of $\pm\sigma_x$ are the same.}
\end{figure}

\begin{figure}[H]
\includegraphics[width=8cm]{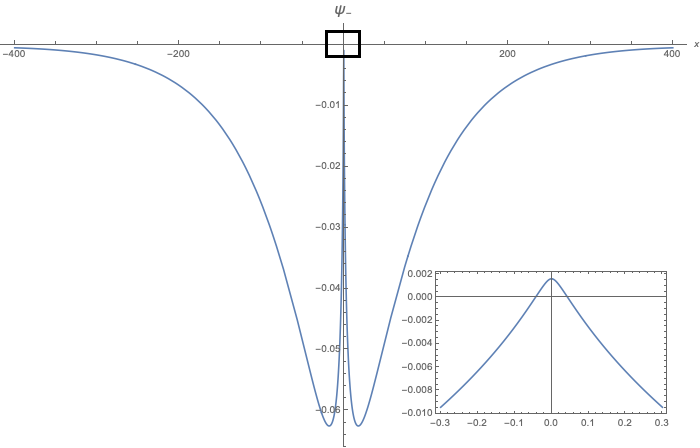}
\includegraphics[width=8cm]{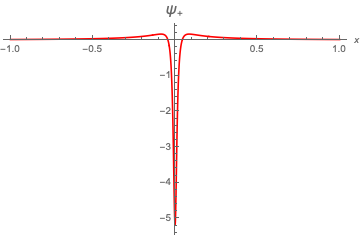}
\centering
\caption{The figure shows the normalized second excited state $\psi_\pm(x)$ generated by numerical integration of the collapse differential equations with $\omega_0=1.3$, $E=-0.500200$ and $\omega=1$. The second excited state wavefunctions have two nodes because of the node theorem. Again, $\int^\infty_{-\infty}\psi^2_+(x)dx=\int^\infty_{-\infty}\psi^2_-(x)dx$.}
\end{figure}

\section{\label{sec:level1}Conclusion}
In this paper, we have developed a tool to study the two-photon Rabi model at the spectral collapse point which is numerically difficult and elusive.  Previously, the incomplete spectral collapse can only be understand qualitatively by numerical method and variational method. Using the real space representation, a pair of second-order differential equations has been derived that enables us to examine the bound states at the collapse point which is a mystery in the previous studies. The spin slitting term $w_0$ is the controlling factor of the depth of the effective potential well, and therefore the number of bound states that can exist at the collapse point are controlled by $\omega_0$. The complete mathematical treatment is still missing but the effective potential gives us a physical intuition and strong evidence about the incomplete spectral collapse. This also explains following numerical observations.
\begin{enumerate}
  \item Spectral collapse happens at $\epsilon=\frac{\omega}{2}$ independent of the value of $\omega_0$.
  \item The bound states of incomplete spectral collapse form only for $E<-\frac{\omega}{2}$.
  \item The energy of ground state is bounded below $E_g>-\omega_0/2$.
  \item The number of bound states increases with the value of $\omega_0$ 
\end{enumerate}
In the previous studies on two-photon Rabi model, only the energy spectrum is exactly obtained using theoretical or numerical method. The explicit wavefunction is still difficult to calculate, but it is necessary for finding many other physical observables. The collapse point differential equations also provide an easier way to generate the wavefunction at the collapse point for further calculation, which is very difficult if other current approximation or numerical methods are used.\\

The two-photon Rabi model can be separated into 3 regimes, before, at and beyond the collapse point $\epsilon=\omega/2$. The effect of $\omega_0$ is well studied in the first two regimes. However, beyond the collapse point, the existence of bound state is still unsolved. A lot of numerical evidences suggest there are no bound state for $\epsilon>\omega/2$ \cite{5,9}. And the theoretical point of view also suggests the asymptotic behaviour of two-photon Rabi model is independent of $\omega_0$, which means all states must be unbounded beyond the collapse point. But still no conclusive work has been done, which needs further investigations.\\

The main reason why this method works is because the system with spectral collapse should exhibit special symmetry at the collapse point. Those special symmetry would simplify the equation, which allows physical intuitive answer. There are some models similar to the two-photon Rabi model. The two-photon Rabi-Stark model also exhibits spectral collapse phenomenon with a different collapse point \cite{38}. The two-mode Rabi model \cite{22}and intensity-dependent Rabi model \cite{24,25} are algebraically equivalent to two-photon Rabi model. They have the same form of equation using the SU(1,1) operators. Similar treatment can also be applied to investigate the property of the bound state at the collapse point for those models. But the real space representation is not trivial for both cases. The two-mode Rabi model involves partial differential equation. The intensity-dependent Rabi model cannot be defined in real space. It is possible to study the collapse property of these models using the method presented in this paper. 

\section{\label{sec:level1}Acknowledgments}
Special thanks to the Department of Physics, the Chinese University of Hong Kong for supporting this research.

\appendix
\section*{Appendix}
\section{Fourier Transform as \texorpdfstring{$\sigma_z$}{} operator in real space}
$\psi_-(x)$ and $\psi_+(x/\omega)$ are the Fourier transform pair. This can be seen by Fourier transform one of the Eq.(\ref{eq:wideeq}). Using the fact 
\begin{equation}
    F[x^n\phi^{(m)}(x)]=(-i)^{(n+m)}\frac{d^n}{dk^n}[k^mF[\phi](k)]
\end{equation}
Here we take the equation for $\psi_+(x)$ and denote $\phi(k)=F[\psi_+(x)]$
\begin{align}
    &\bigg(1-\frac{4\epsilon^2}{\omega^2}\bigg) F[\psi''''_+(x)]-2\bigg(1+\frac{4\epsilon^2}{\omega^2}\bigg)\omega^2F[x^2\psi''_+(x)] -2(2E+\omega) F[\psi''_+(x)]
-4\bigg(1\pm\frac{2\epsilon}{\omega}\bigg)^2\omega^2F[x\psi'_+(x)]\nonumber\\
&+\bigg(1-\frac{4\epsilon^2}{\omega^2}\bigg)\omega^4F[x^4\psi_+(x)]-(4E+2\omega)\omega^2F[x^2\psi_+(x)]+\bigg[(2E+\omega)^2-2\bigg(1\pm\frac{2\epsilon}{\omega}\bigg)^2\omega^2-\omega_0^2\bigg]F[\psi_+(x)]=0\nonumber\\
    &\bigg(1-\frac{4\epsilon^2}{\omega^2}\bigg) k^4\phi(k)-2\bigg(1+\frac{4\epsilon^2}{\omega^2}\bigg)\omega^2[k^2\phi(k)]'' -2(2E+\omega) k^2\phi(k)
-4\bigg(1\pm\frac{2\epsilon}{\omega}\bigg)^2\omega^2[k\phi(k)]'\nonumber\\
&+\bigg(1-\frac{4\epsilon^2}{\omega^2}\bigg)\omega^4\phi''''(k)-(4E+2\omega)\omega^2\phi''(k)+\bigg[(2E+\omega)^2-2\bigg(1\pm\frac{2\epsilon}{\omega}\bigg)^2\omega^2-\omega_0^2\bigg]\phi(k)=0\nonumber
\end{align}

After arranging the above equation, we can get
\begin{align}
\bigg(1-&\frac{4\epsilon^2}{\omega^2}\bigg)\omega^4 \phi''''(k)-\bigg[2\bigg(1+\frac{4\epsilon^2}{\omega^2}\bigg)\omega^2k^2-2(2E+2\omega)\omega^2\bigg]\phi''(k) 
-4\bigg(1-\frac{2\epsilon}{\omega}\bigg)^2\omega^2k\phi'(k)\nonumber\\
&+\bigg[\bigg(1-\frac{4\epsilon^2}{\omega^2}\bigg)k^4-(4E+2\omega)k^2+(2E+\omega)^2-2\bigg(1-\frac{2\epsilon}{\omega}\bigg)^2\omega^2-\omega_0^2\bigg]\phi(k)=0
\label{ft}
\end{align}

Then we can substitute $x=\omega k$ in Eq.(\ref{ft}). We can recover Eq.(\ref{eq:wideeq}) of $\psi_-(x)$ which will equal to squeezed Fourier transform of $\psi_+(x)$. \\

In fact, this property can be explained physically. The Fourier transform in real space is actually equivalent to the spin $\sigma_z$ operator in two-photon Rabi model. The logic is as follow. It is well known that the two-photon Rabi model can separated into 4 different sectors, formed by even or odd number states and spin-up or spin-down paired with the first number state. Consider the even and spin down sector, the wavefunction can be decomposed as superposition of even number states with alternating spin
\begin{equation}
    \ket{\Psi}=\sum_n a_{4n}\ket{4n,\downarrow}+\sum_n a_{4n+2} \ket{4n+2,\uparrow}
\end{equation}
The number states (of $\omega=1$) are eigenstates of Fourier transform with eigenvalues $(-i)^n$
\begin{equation}
    F[\ket{n}]=(-i)^n\ket{n}
\end{equation}
For example, $\ket{0}$ is just a Gaussian function which is invariant under Fourier transform. When $\ket{\Psi}$ is taking Fourier transform, it is just Fourier transform each $\ket{n}$ basis. 
\begin{align}
    F[\ket{\Psi}]&=\sum_n a_{4n}F[\ket{4n,\downarrow}]+\sum a_{4n+2}F[\ket{4n+2,\uparrow}]\nonumber\\
                &=\sum_n a_{4n}i^{4n}\ket{4n,\downarrow}+\sum a_{4n+2}i^{4n+2} \ket{4n+2,\uparrow}\nonumber\\
                &=\sum_n a_{4n}\ket{4n,\downarrow}-\sum a_{4n+2} \ket{4n+2,\uparrow}\nonumber\\
                &=-\hat{\sigma}_z\Psi
\end{align}
Because of the parity conservation, the eigenstates of two-photon Rabi model can be decomposed into superposition of even or odd number states with alternating spin. The Fourier transform acting on the number state is actually the same as acting $i^c\sigma_z$ on the spin part where $c=\{0,1,2,3\}$ depends on the sector of the eigenstate.\\

Using Eq.(\ref{eq:1}) and the Fourier transform relation of $\psi_\pm(x)$ ($\omega=1$), we can derive another equation for $\psi_\pm(x)$ at the collapse point,
\begin{equation}
    \omega_0F[\psi_-(k)]=(-2x^2+2E+1)\psi_-(x)
    \label{Fn}
\end{equation}
\begin{equation}
    \omega_0F[\psi_+(k)]=\bigg(2\frac{d^2}{dx^2}+2E+1\bigg)\psi_+(x)
    \label{Fp}
\end{equation}

These equations show a special property of $\psi_\pm(x)$ where their Fourier transform have a similar form of themselves. These equations are mathematically interesting because Eq.(\ref{Fn}) and (\ref{Fp}) are eigenvalue equations involving Fourier transform. The solution of these equations are also not yet studied. But this provides an alternative way to study the problem mathematically.
\printbibliography

\end{document}